\documentclass[3p]{elsarticle}

 \usepackage{graphicx}
\usepackage{lscape} 
\usepackage{rotating}

\usepackage{amssymb}

\usepackage{units}

\usepackage{amsmath}
\def\Journal#1#2#3#4{ {#1} {\bf #2}, #3 (#4)}

\def\MPA{{\em Mod. Phys. Lett.} A}

\def\NPA{{\em Nucl. Phys.} A}
\def\PLB{{\em Phys. Lett.} B}

\def\PRL{\em Phys. Rev. Lett.}
\def\PRD{{\em Phys. Rev.} D}
\def\PRC{{\em Phys. Rev.} C}

\def\PR{{\em Phys. Rev.}}
\def\PRP{{\em Phys. Rep.}}

\def\ZPA{{\em Z. Phys.} A}

\def\PAN{\em Phys. Atom. Nucl.}

\def\ARI{\em Appl. Rad. and Isotopes}

\def\EPL{\em Europhys. Lett.}

\def\EPJA{{\em Eur. Phys. J.} A}

\def\ra{\rightarrow}
\def\be{\begin{equation}}
\def\ee{\end{equation}}
\def\bea{\begin{eqnarray}}
\def\eea{\end{eqnarray}}
\def\beas{\begin{eqnarray*}}
\def\eeas{\end{eqnarray*}}

\newcommand{\obb}{0\mbox{$\nu\beta\beta$-decay}}
\newcommand{\zbb}{2\mbox{$\nu\beta\beta$-decay}}

\newcommand{\pdhz}{\mbox{$^{110}$Pd}~}
\newcommand{\pdht}{\mbox{$^{102}$Pd}}

\newcommand{\nel}{\mbox{$\nu_e$}}
\newcommand{\bpbp}{\ensuremath{\beta^+\beta^+}~}
\newcommand{\bec}{\ensuremath{\beta^+/{\rm EC}}}
\newcommand{\ecec}{\ensuremath{{\rm EC/EC}}~}
\newcommand{\ema}{\ensuremath{\langle m_{\nu_e} \rangle}~}

\begin{document}

\begin{frontmatter}




\title{A first search of excited states double beta and double electron capture decays of $^{110}$Pd and $^{102}$Pd}

\author[dresden]{B. Lehnert\corref{cor1}}
\ead{Bjoern.Lehnert@mailbox.tu-dresden.de}
\author[dresden]{K. Zuber}
\ead{Zuber@physik.tu-dresden.de}
\address[dresden]{Inst. f\"ur Kern- und Teilchenphysik, Technische Universit\"at Dresden, 01069 Dresden, Germany}

\cortext[cor1]{Corresponding author}

\author{}

\address{}

\begin{abstract}
A search for double beta decays of the palladium isotopes \pdhz and \pdht\ into excited states of their daughters was performed and first half-life limits for the $2\nu\beta\beta$ and $0\nu\beta\beta$ decays into first excited 0$^+$ and $2^+$ states of \unit[$5.89 \times 10^{19}$]{yr} and \unit[$4.40 \times 10^{19}$]{yr} (\unit[95]{\%} CL) for the \pdhz decay were obtained. The half-life limits for the corresponding double electron capture transition of \pdht\ are \unit[$7.64 \times 10^{18}$]{yr} and \unit[$2.68 \times 10^{18}$]{yr} (\unit[95]{\%} CL) respectively. These are the first measurements for \pdht.
\end{abstract}

\end{frontmatter}

\section{Introduction}
\label{intro}
During the past 20 years vast progress has been made in unveiling the 
properties of neutrinos. For decades neutrinos were thought to be massless, 
which no longer holds true: flavour oscillations found in the leptonic sector, studying neutrinos 
coming from the sun \cite{sno,superKsolar}, the atmosphere \cite{superKatmos}, high energy 
accelerators beams \cite{minos10,opera10} and nuclear power plants \cite{kamland03}, are explained by neutrino oscillations requiring a non zero neutrino mass.
However, no absolute mass scale can be fixed with experiments studying the oscillatory behaviour. 
To achieve this, one has to investigate weak decays, such as beta decays or neutrinoless double beta decays

\be
(Z,A) \ra (Z+2,A) + 2 e^-  \quad (\obb) \, . 
\ee

The latter violates total lepton number by two units and thus is not allowed in the Standard Model. The \obb\ is the gold plated process to distinguish whether neutrinos are Majorana or  Dirac particles. Furthermore, a match of helicities of the intermediate neutrino states is necessary which is done in the easiest way by introducing a neutrino mass. This mass is linked with the experimentally observable half-life via

\begin{equation}
  \label{eq:1}
 \left(T_{1/2}^{0 \nu}\right)^{-1} = G^{0 \nu}(Q, Z) \left| M_{GT}^{0\nu} - M_{F}^{0\nu} \right|^2 \left(\frac{\ema}{m_e}\right)^2 \, ,
\end{equation}

where \ema is the effective Majorana neutrino mass, given by $\ema = \left| \sum_{i}U_{ei}^2m_i\right|$ and $U_{ei}$ is the corresponding  element in the 
leptonic PMNS mixing matrix, $G^{0 \nu}(Q, Z) $ is a phase space factor and $M_{GT}^{0\nu} - M_{F}^{0\nu}$ describes the nuclear transition matrix element. 
The experimental signature is the emission of two electrons with a sum energy 
corresponding to the Q-value of the nuclear transition. 
A potential evidence has been claimed in the  \obb\ of  $^{76}$Ge with $T_{1/2}^{0 \nu} = \unit[2.23^{+0.44}_{-0.31} \times 10^{25}]{yr}$ at \unit[90]{\%} CL \cite{kla06}. In addition, the SM process of neutrino accompanied double beta decay,

\be
(Z,A) \ra (Z+2,A) + 2 e^- + 2 \nel \quad (\zbb) \,
\ee

can be investigated, which is expected with half-lives arround \unit[$10^{20}$]{yr}. For recent reviews see \cite{avi08}.

Additional information is provided by the alternative process of positron decay in combination with electron capture (EC). 
Three different decay modes can be considered:

\begin{align}
(Z,A) \ra& (Z-2,A) + 2 e^+ + (2 \nel)  &\mbox{(\bpbp{})}\\
e^- + (Z,A) \ra& (Z-2,A) + e^+ + (2 \nel) & \mbox{(\bec{})}\\
2 e^- + (Z,A) \ra& (Z-2,A) + (2 \nel)  &\mbox{(\ecec{})}
\end{align}

Decay modes containing a positron have a reduced Q-value as each generated positron accounts for a reduction 
of 2 $m_ec^2$. Thus, the full energy is only available in the \ecec mode and makes it the most probable one.
However, it is also the most difficult to detect, only producing X-rays instead of \unit[511]{keV} gammas.
Furthermore, it has been shown that \bec\ transitions have an enhanced sensitivity to right-handed weak
currents (V+A interactions) \cite{hir94} and thus would help to disentangle the physics mechanism of \obb.
In the last years, also neutrinoless \ecec modes have been discussed with renewed interest, because of a potential
resonance enhancement up to a factor of $10^6$ in the decay if the initial and final excited state are degenerate 
\cite{suj04}. Recently, a series of isotopes with extreme low Q-value was explored for enhancement in the ECEC mode
and with $^{152}$Gd a very promising candidate was found \cite{eli11}.

Another branch of search is linked to excited state transitions. The signal in this case is extended by looking at
the corresponding de-excitation gammas. However, in the approach of a passive sample on a Ge-detector
it will not allow to distinguish between the \zbb\ and \obb\ mode. Thus, the deduced half-live limits are valid for both.
The investigation of \zbb\ modes into excited states will add information on nuclear structure, valuable for matrix
element calculations. Furthermore, a potential observation of \obb\ into an excited $2^+$-state would likely point to other
contributions besides neutrino masses. The searches described in this paper are based on the search for excited
state transitions.

An element  getting little attention in the past is palladium with the isotopes of interest \pdhz and \pdht. 
Among the eleven \obb\ candidates with a Q-value larger than \unit[2]{MeV}, \pdhz has several advantages: it has the second-highest natural abundance (\unit[11.72]{\%}) and, in addition, it is an excellent candidate to probe the single-state dominance hypothesis for \zbb, i.e. that only the lowest lying intermediate 1$^+$-state will contribute to the nuclear transition matrix element describing its \zbb. Only one rather weak experimental limit in the order of \unit[$10^{17}$]{yr} exists for \obb\ ground state transitions in \pdhz \cite{win52}. Theoretical predictions for the \zbb\ ground state mode are in the range of \unit[0.12 - 29.96 $\times$ 10$^{20}$]{yr} \cite{chandra05,semenov2000,civitarese1998,stoica1994,hirsch1994,staudt1990,domin2005,suhonen2011}. Theoretical predictions for the excited state transitions are \mbox{\unit[4.4 $\times$ 10$^{25}$]{yr}} \cite{domin2005}, \mbox{\unit[8.37 $\times$ 10$^{25}$]{yr}} \cite{stoica1994}, \mbox{\unit[1.5 $\times$ 10$^{25}$]{yr}} \cite{raduta2007} and \mbox{\unit[0.62 - 1.3 $\times$ 10$^{25}$]{yr}} \cite{suhonen2011} for the $2^+_1$ state and \mbox{\unit[2.4 $\times$ 10$^{26}$]{yr}} \cite{domin2005} and \mbox{\unit[4.2 - 9.1 $\times$ 10$^{23}$]{yr}} \cite{suhonen2011} for the $0^+_1$ state.

\begin{figure}[h]
\centering
\includegraphics[width=16cm]{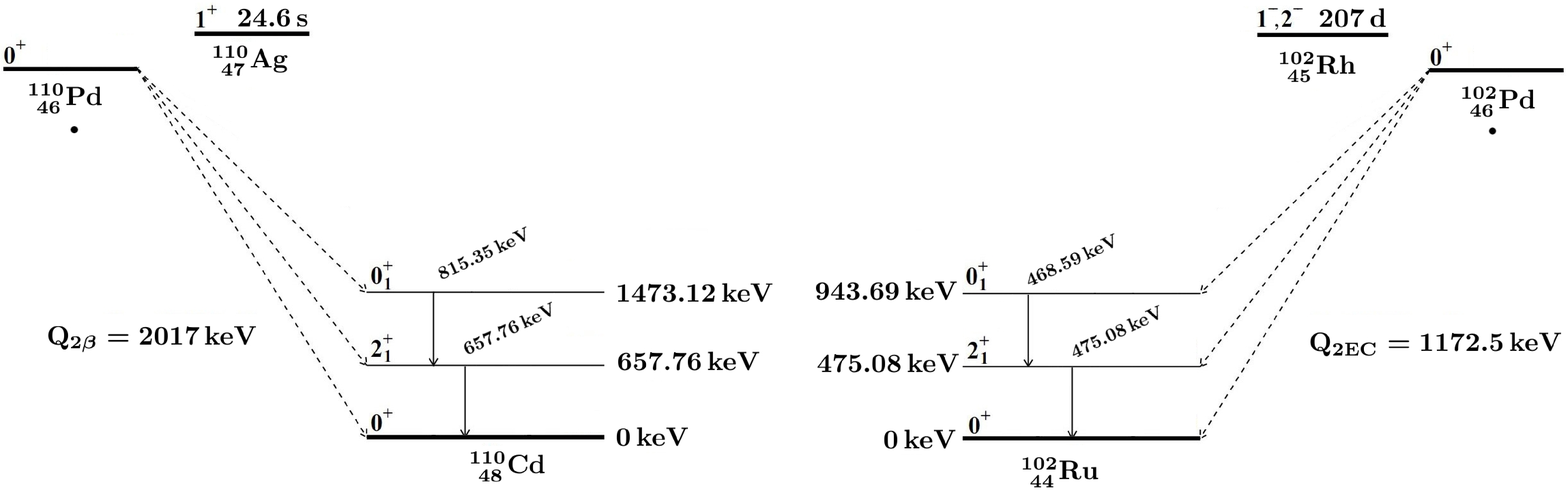}
\caption{Level schemes of  \pdhz (left) and \pdht\ (right) decays.}
\label{fig:levels}
\end{figure}

The second isotope \pdht\ has a Q-value of \unit[1172]{keV}, a natural abundance of \unit[1.02]{\%} and is able to decay via \ecec and 
\bec. These decay modes have never been studied for \pdht\ experimentally and no theoretical predictions exist. The level schemes
of both isotopes are shown in Fig.~\ref{fig:levels}.

\section{Experimental Setup}
\label{setup}
The measurement was performed in the Felsenkeller Underground Laboratory in Dresden with a shielding depth of 120 mwe. 
A sample of \unit[802.35]{g} of Pd was used, which was purified before by C. HAFNER GmbH + Co. KG. 
It was placed in a standard Marinelli baker (D6) of \unit[70]{cm}
diameter and \unit[21]{cm} height. The sample was positioned on a HPGe detector with an efficiency of \unit[90]{\%} routinely used for $\gamma$-spectroscopic measurements. The detector has an \unit[1]{mm} thick Al-window towards the sample.
It is surrounded by a \unit[5]{cm} copper shielding embedded in another shielding of \unit[15]{cm} of clean
lead. The inner \unit[5]{cm} of the lead shielding have a low contamination of $^{210}$Pb of only \unit[$2.7 \pm 0.6$]{Bq/kg} while
the outer \unit[10]{cm} have an activity of \unit[$33 \pm 0.4$]{Bq/kg}.

\begin{figure}[h]
\begin{center}
  \includegraphics[width=10cm]{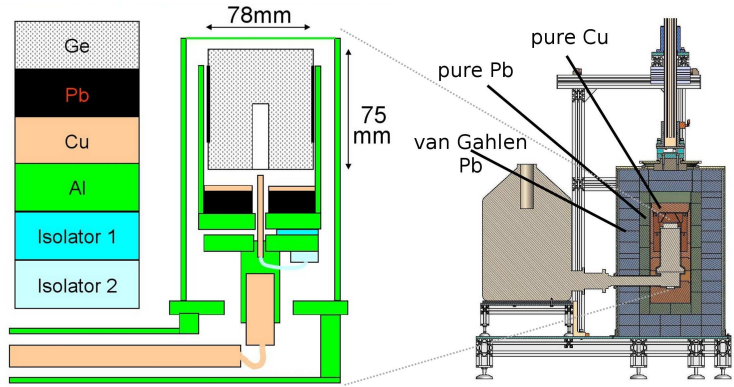}

\caption{\small Schematic drawing of the used setup. Composed from pictures in \cite{degering08}}
\label{fig:schema}
\end{center}
\end{figure}

The detector is placed inside a special building which acts as a Faraday cage and a Rn shield; additionally, the setup is flushed with nitrogen in order to reduce Rn contamination. More details can be found in \cite{deg09,degering08} and a schematic drawing is shown in Fig.~\ref{fig:schema}. The Pd was stored underground for more than one year prior to the measurement except for \unit[18]{d} of purification. The data were collected with a 8192 channel MCA from Ortec and were converted into the ROOT format for analysis.

An extensive calibration was performed using 17 $\gamma$-lines from 8 different nuclides resulting in a linear energy calibration curve of

\be
E [\unit{keV}] = \unit[0.342746]{\frac{\unit{keV}}{channel}\times channel} - \unit[4.337734]{keV}\, .
\ee

The measurement range of the spectrum goes up to \unit[2.8]{MeV}.
The resolution was calibrated using the calibration lines and fitted with a second order polynomial.
The actual values at the energies of the lines of interest will be discussed in the corresponding analysis section.

Despite purification, the measured spectrum is dominated by intrinsic contaminations of the Pd. Clear $\gamma$-lines from the $^{238}$U and $^{232}$Th
decay chains as well as $^{40}$K are visible. However, former Americium contributions have been removed completely by
the purification.
The actual background spectrum of the detector system itself without any sample is at least an order of magnitude smaller
in the regions of interest and can be neglected, hence the spectrum is completely dominated by the Pd sample
contaminations.


\section{Analysis and Results}
A total of \unit[16.2]{d} of data were accumulated resulting in \unit[13.00]{$\rm kg \cdot d$}
of exposure. 
In the following the two isotopes of interest are discussed separately.
For the analysis, the natural abundances of the latest IUPAC evaluation
have been used which are \unit[11.27]{\%} (\pdhz) and \unit[1.02]{\%} (\pdht) respectively \cite{bie93}.
As the searches are purely based on gamma detection, the obtained results
apply for both, \obb\ and \zbb\ modes. The decays into the first excited $0_1^+$-state
de-excite via an intermediate $2^+_1$- state. Thus, there will be an angular correlation
among the gammas, with an observational probability W($\theta$) that the second 
gamma is emitted with an angle $\theta$ with respect to the first one given by

\be
W(\theta)= \frac{5}{4} \left(1 - 3 \cos^2 \theta +4 \cos^4 \theta\right)\, .
\ee

It can be seen that the probability of both gammas beeing emitted in the same direction is larger than an uncorrelated emission. However, the detection efficiency for a single gamma is small which results in a low probability to observe a summation peak and thus the searches are based on the individual gamma energies only.

\begin{figure}[h]
\begin{center}
  \includegraphics[width=10cm]{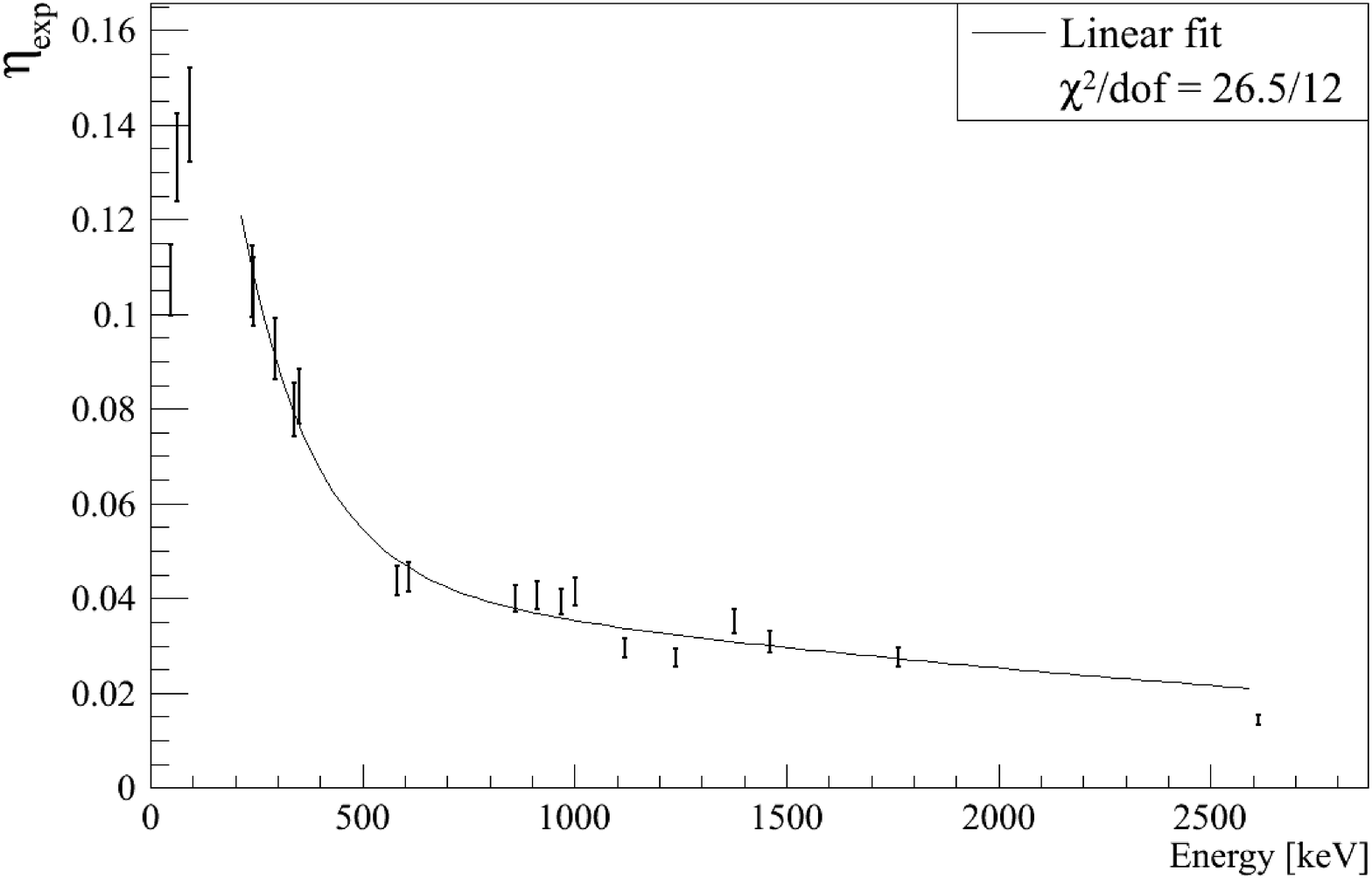}
\caption{\small Measured efficiencies and the fitting function.}
\label{fig:efficency}
\end{center}
\end{figure}

The efficiency for full energy detection was determined and cross checked in several ways.
The most important one was replacing the actual used volume by a SiO$_2$ sample
of exactly the same geometry. The intrinsic contaminations of the natural decay chains of $^{238}$U
and $^{232}$Th as well as $^{40}$K produced various $\gamma$-lines and acted as an extended calibration source. The 17 $\gamma$-lines used for the energy calibration were also used for the efficiency determination
in the region from \unit[238.6]{keV} (from $^{212}$Pb decay) up to \unit[2614.3]{keV} (from $^{208}$Tl decay). 
The efficiency in the region above \unit[200]{keV} can be well fitted by two exponential functions.
It varies between \unit[5.77]{\%} at \unit[468]{keV} and \unit[3.89]{\%} at \unit[815]{keV}, being the lowest and highest
energy lines of interest for the search described in this paper.
To account for the difference in self-absorption of Pd
and SiO$_2$, extensive Monte Carlo simulations were performed using the AMOS code 
\cite{amos}. The amount of Monte Carlo was chosen in a way that the statistical error for the
full energy peak for both, Pd and SiO$_2$, was less than \unit[0.1]{\%}.
The simulations agree within an error of less than \unit[15]{\%} with the measurements and tend to
be slightly higher. This can easily be explained by small geometric differences in the simulation
and the experiment. However, independent from that is the ratio of both self-absorption simulations for Pd and SiO$_2$.
Hence, the ratio was used to scale the experimentally well determined efficiency curve of SiO$_2$ to the one of Pd (Fig.~\ref{fig:efficency}). 
A validation of the procedure was performed within the vicinity of the lines of interest
by using prominent background lines apparent in the spectrum, 
namely \unit[295.21]{keV} and  \unit[351.92]{keV} (from $^{214}$Pb), 
\unit[233.63]{keV}  (from $^{212}$Pb), \unit[583.19]{keV}  (from $^{208}$Tl) and \unit[609.32]{keV}  (from $^{214}$Bi),
which results in good agreement.

The analysis is based on the extraction of upper count limits with the Feldman Cousins method described in \cite{fc98}. A simple constant was chosen as a background model after concluding that the area around the ROI is sufficiently flat. The background was determined with a likelihood fit in a selected region around the peak position taken to be \unit[$\pm 30$]{keV} excluding the peak range which is considered to be \unit[$\pm 5$]{keV}. 

\begin{figure}[H]
\begin{center}
 \includegraphics[width=10cm]{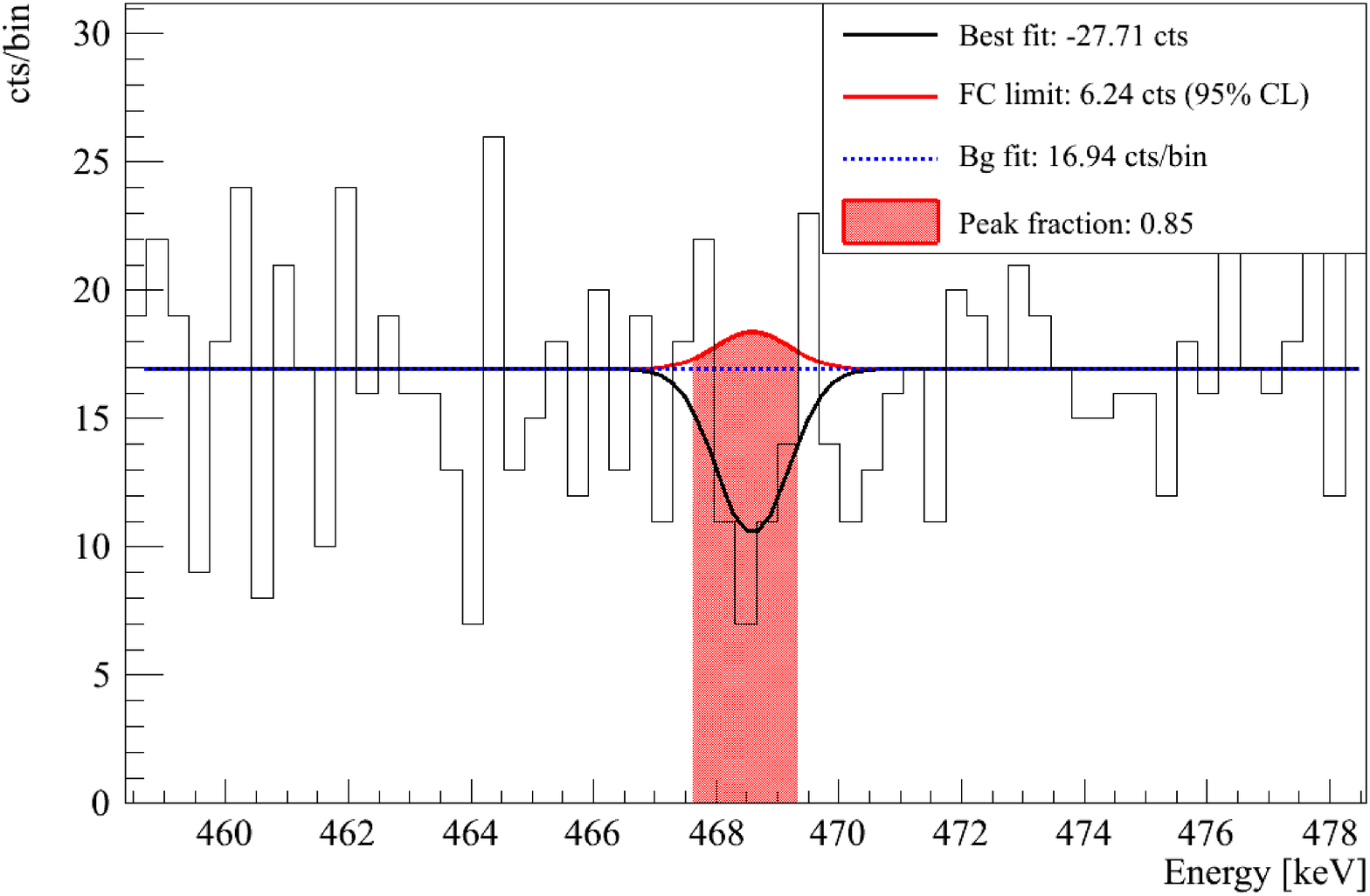}
\caption{\small Peak region of \pdht\ for the $0^+ \rightarrow 2^+$ transition with the illustration of the background model, the best fit and the peak with the upper count limit at \unit[95]{\%} CL.}
\label{fig:pd102}
\end{center}
\end{figure}

\begin{figure}[H]
\begin{center}
 \includegraphics[width=10cm]{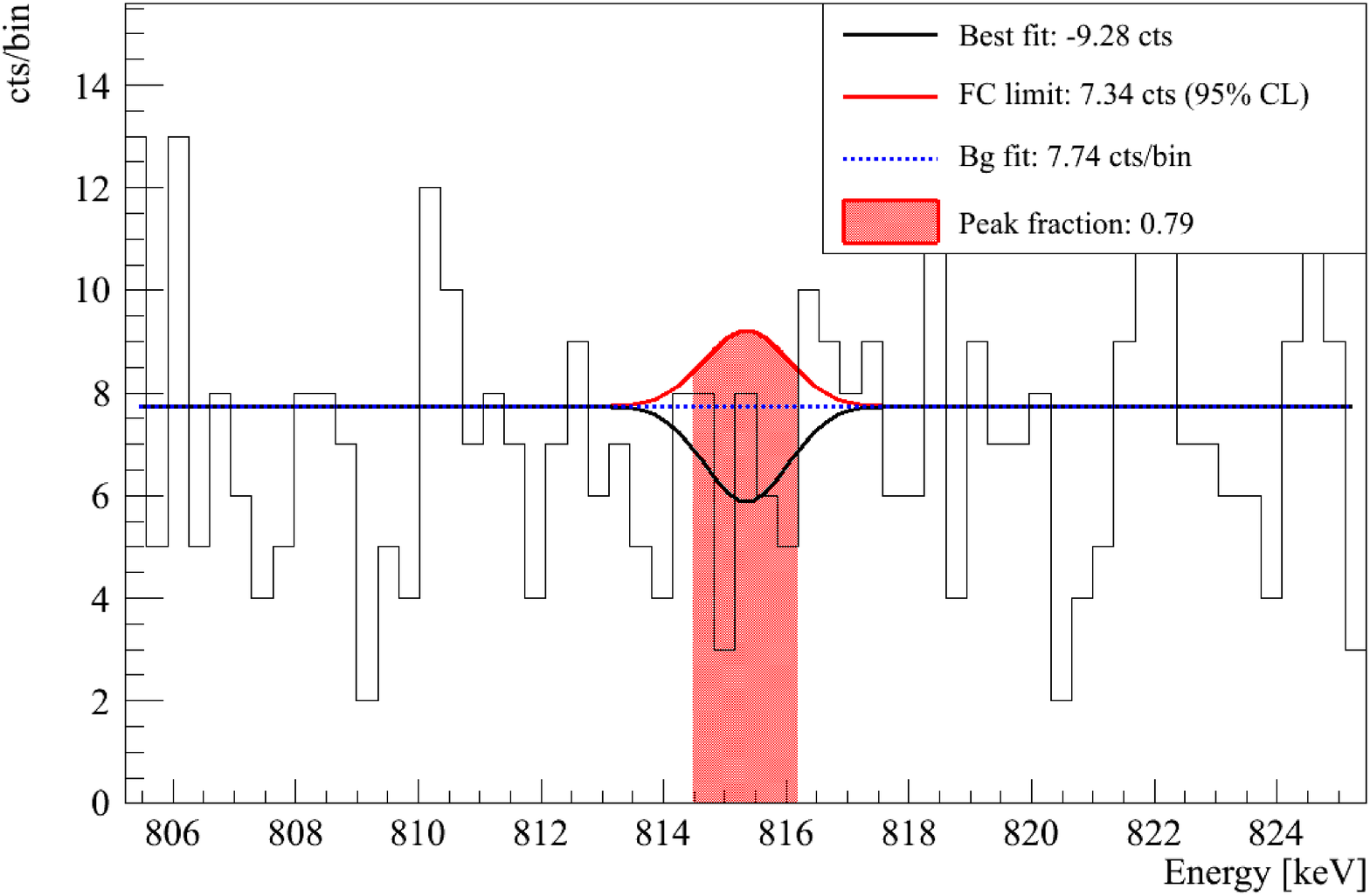}
\caption{\small Peak region of \pdhz for the $0^+ \rightarrow 2^+$ transition with the illustration of the background model, the best fit and the peak with the upper count limit at \unit[95]{\%} CL.}
\label{fig:pd110}
\end{center}
\end{figure}

Likelihood fits with the Gaussian peak shapes and the background constants resulted in statistical downward fluctuations for all four peaks. This is a clear sign that no significant signal is observed (see Fig.~\ref{fig:pd102} and \ref{fig:pd110}). In this case, an upper limit for the count rate can be calculated using the background only hypothesis and the maximal statistical fluctuation of a Gaussian distributed background for a certain confidence level. This is commonly referred to as sensitivity. However, the Feldman Cousins approach also considers the observed downward fluctuations and results in more appropriate results for low count rates.

In order to obtain a numerical value for the upper limit, all bins within the FWHM of an expected peak are combined into a single analysis bin. The resulting fraction of the peak that is covered by the analysis bin is dependent on the binning of the data but always larger than \unit[76]{\%}. The background expectation and the measured count rate are used to evaluate an upper count limit for this bin with the ROOT implementation of the Feldman and Cousins method. The result is then scaled to the full peak area.

From the background point of view, exactly the same $\gamma$-lines in the performed search could be produced from the beta decays of the 
intermediate nuclide of the investigated double beta system, which will be discussed in more detail in the next
section. However, their contribution can be rejected by the non-observation of other, 
more prominent $\gamma$-lines at different locations in the spectrum.
The only prominent background line to be expected within \unit[$\pm 5$]{keV} of any of the four
lines under investigation is from $^{137}$Cs at \unit[661.66]{keV} potentially influencing the $2^+_1$-limit in the \pdhz system.
However, no indication of this line is observed.

%
\subsection{The \pdhz system}
\label{data}
Two lines were investigated for \pdhz at energies \unit[657.76]{keV} (corresponding to the 2$^+_1$ transition)
and \unit[815.35]{keV} (additionally emitted in the $0^+_1$ decay) respectively.
The corresponding energy resolutions at these energies are 1.51 and \unit[1.61]{keV} (FWHM).
Potential $\gamma$-lines mimicking the signal would come from $^{110}$Ag and $^{110m}$Ag
decays. $^{110}$Ag has a half-life of 24.6 s only and thus has to be produced in-situ. With the
given shielding this can be excluded. Potentially more dangerous is the long-living $^{110m}$Ag
(half-live of 249.79 d). This isotope has two prominent lines at \unit[1384.3]{keV} and \unit[1505.04]{keV}. They are not visible in the spectrum and thus this contribution can be excluded for this search.
No lines were visible at both peak positions of interest and thus an upper limit (\unit[95]{\%} CL) 
of signal events of 10.53 and 7.34 could be extracted for \unit[657.76]{keV} and \unit[815.35]{keV} respectively.

Using the known Pd mass and efficiencies, this can be converted into lower half-live limits of 

\begin{align}
T_{1/2}^{(0\nu + 2\nu)}\, \pdhz \ra \rm {^{110}Cd} (0^+_1,\unit[815.3]{keV}) > \unit[5.89 \times 10^{19}]{yr}\ (\unit[95]{\%} CL)& \\
T_{1/2}^{(0\nu + 2\nu)}\, \pdhz \ra \rm {^{110}Cd} (2^+_1,\unit[657.8]{keV}) > \unit[4.40 \times 10^{19}]{yr}\ (\unit[95]{\%} CL)& \, .
\end{align}

These are the first experimental limits for excited state transitions in the \pdhz system.

\subsection{The \pdht\ system}
Two lines were investigated for \pdht\ at energies \unit[468.59]{keV} (only emitted in the $0^+_1$ decay) and 
\unit[475.05]{keV} (corresponding to the 2$^+_1$ transition) respectively.
The energy resolution at these energies is \unit[1.39]{keV} (FWHM) for both lines.
Potential $\gamma$-lines mimicking the signal would come from $^{102}$Rh and $^{102m}$Rh
decays. $^{102}$Rh with a half-life of 207 d has no reasonable line to check. The strongest one is the \unit[475.05]{keV} line.
As there is no signal in this region it can be concluded that it does not contribute to the \unit[468.59]{keV}
region. On the other hand $^{102m}$Rh (half-live of \unit[2.9]{yr}) has multiple lines to explore, the most restricting ones
are a line at \unit[631.28]{keV} with \unit[56]{\%} emission probability and at \unit[697.49]{keV} with \unit[44]{\%}. Both of them
are not observed in the spectrum and thus can exclude such a contribution.
No lines were visible at both peak positions of interest and thus an upper limit (\unit[95]{\%} CL) 
of signal events of 17.64 and 6.24 could be extracted for \unit[475.05]{keV} and \unit[468.59]{keV} respectively. 
The obtained half-live limits are 

\begin{align}
T_{1/2}^{(0\nu + 2\nu)}\, \pdht\ \ra \rm {^{102}Ru} (0^+_1,\unit[468.6]{keV}) > \unit[7.64 \times 10^{18}]{yr}\ (\unit[95]{\%} CL)&  \\
T_{1/2}^{(0\nu + 2\nu)}\, \pdht\ \ra \rm {^{102}Ru} (2^+_1,\unit[475.1]{keV}) > \unit[2.68 \times 10^{18}]{yr}\ (\unit[95]{\%} CL)& \, .
\end{align}

These are the first experimental limits on \pdht\ double beta decays.

\section{Summary}
\label{conclusion}
Double beta decay transitions into excited states for the two Pd-isotopes \pdht\ and \pdhz have been
investigated for the first time. These transitions contain valuable informations about the physics mechanism of double beta
decay and the involved nuclear physics. However, no signal into the first excited 0$^+$ and 2$^+$ states have
been observed.

\section*{Acknowledgement}
The authors would like to thank D. Degering (VKTA Dresden) for his help with the
underground measurements and D. Sommer for her help with the Monte Carlo
simulations for efficiency determinations.

%

\end{document}